# A Simple Viscometer for High School and First Years Undergraduate Program Students: Theory and Experiment

Sparisoma Viridi[1,*], Sidik Permana[1], Wahyu Srigutomo[1], Anggie Susilawati[2], and Acep Purqon[1]

[1]Physics Department, Institut Teknologi Bandung, Bandung 40132, Indonesia
[2]Physics Department, Universitas Padjajaran, Sumedang 45363, Indonesia
Email: dudung@fi.itb.ac.id

*Abstract*

*A simple viscometer consists of mineral water bottle and trink straw is used to predict vicosity of water. A model is derived from Bernoulli's principle and Poiseulli's law. Value about 0.374 – 0.707 cP is obtained for observation in room temperature. About 10 minutes of observation time is needed to get the data.*

## Introduction

In discussing fluid flow, equation of continuity and Bernoulli’s equation are the most common discussed [1], while Poiseulli flow for real fluid with viscosity [2], that produces formula relating flow rate and pressure difference through a pipe with certain length and radius [3] is rarely mentioned. In some daily life problem such as leakage on a water container, the latest concept is needed, since the two first concepts predict time for the container to be drained out is the same for all fluids, through formula known as Torricelli law [4, 5]. It seems that the drained out time is independent to fluid viscosity coefficient, which is incorrect. A common method to determine fluid viscosity coefficient is using a ball falling in viscous fluid [6], which can be extended to buoyant ball experiment [7]. In this work the Torricelli law will be extended using Poiseulli flow to include influence of fluid viscosity coefficient and a robust experiment is performed to justify the formulation in determining fluid viscosity coefficient. Influence of temperature, such as in an empirical formula [8], to fluid visocisity coefficient is neglected in this work.

Required theories for building a simple viscometer and sketch of the system are explained briefly in theory part. Daily small things, that can be found easily at home, are components of the viscometer. They will be listed in experiment part. The next two parts are results and discussion part and conclusion part.

## Theory

Three concepts are used in this work, equation of continuity, Bernoulli’s equation, and Poiseulli flow, where each of them is brief reviewed and illustrated as in the following subsections. In these subsections index $i$ means inlet, while index $o$ means outlet.

### *Equation of continuity*

Through a circular pipe with inlet radius  and outlet radius  a relation between inlet velocity and outlet velocity  is known as equation of continuity

$$v_i A_i = v_o A_o \, . \tag{1}$$

with meaings of $A_i = \pi R_i^{\;2}$ and $A_o = \pi R_o^{\;2}$ as shown in Figure 1. The term debit

$$Q = vA \tag{2}$$

is also common used in Equation (1).

### *Bernoulli’s equation*

Bernoulli’s equation is derived from theorem of work – kinetic energy, which considers only works done by external pressure and gravitation. The equation is

$$p_i + \frac{1}{2}\rho v_i^{\;2} + \rho g y_i = p_o + \frac{1}{2}\rho v_o^{\;2} + \rho g y_o \, , \tag{3}$$

with $\rho$ is fluid density, $p$ is pressure, and $y$ is vertical position of pipe part as illustrated in Figure 2.

Equation of continuity in Equation (1) or (2) and Bernoulli’s equation in Equation (3) are derived for non-compressible fluid. And in its derivation, the last also neglected fluid friction.

### *Poiseulli flow*

In a horizontal circular pipe with length $L$ and radius $R$ , fluid with viscosity coefficient $\eta$ can flow with flow rate $Q$ due to pressure difference $\Delta p$ that follows [3]

$$\Delta p = \frac{8\eta L}{\pi R^4} Q \, . \tag{4}$$

as illustrated in Figure 3.

*System*

A fluid container with form of a cylinder with radius is placed standing that its axis is parallel to direction of gravitation. Near bottom of the container small pipe with radius and length is attached perpendicular to container axis. Distance between fluid surface in the container and position of the small pipe is defined as . The system is illustrated in Figure 4.

Using Equation (3) the relation between inlet point and $Z$ point can be written as

$$P_A + \frac{1}{2}\rho v_i^{\,2} + \rho g h = p_Z + \frac{1}{2}\rho v_Z^{\,2}, \quad (5)$$

with $p_i = P_A$, where $P_A$ is air atmospheric pressure.

It can be assumed that $v_Z \approx 0$ since fluid only flow from inlet to outlet and only disturbs point $Z$ slightly. Then it can traight forward found that

$$p_Z = P_A + \rho g h + \frac{1}{2}\rho v_i^{\,2} \approx P_A + \rho g h, \quad (6)$$

if it is also assume that $v_i$ small and then $v_i^{\,2} \approx 0$.

Height change of water surface $h$ is

$$-\frac{dh}{dt} = v_i. \quad (7)$$

Then substitute Equation (1) and (2) into Equation (7) will give

$$\frac{dh}{dt} = -\frac{Q_0}{A_i}. \quad (8)$$

Equation (4) can also be written for outlet as

$$p_Z - P_A = \frac{8\eta L_o}{\pi R_o^{\,4}} Q_o. \quad (9)$$

with $p_o = P_A$. Substitution Equation (6) into Equation (9) and the result into Equation (8) will give a first order differential equation

$$\frac{dh}{dt} + \frac{\pi R_o^{\,4}\rho g h}{8\eta L_o A_i} = 0, \quad (10)$$

which has solution

$$h(t) = h_o \exp\left[-\left(\frac{\pi R_o^{\,4}\rho g}{8\eta L_o A_i}\right)t\right]. \quad (11)$$

## Experiment

Value of parameters in experiment are $R_i = 4.25$ cm, $R_o = 1.75$ mm, $L_0 = 10$ cm, $g = 9.8$ m/s$^2$, $\rho = 1000$ kg/m$^3$, and $h_0 = 22$ cm. Components of the simple viscometer and after they assambled are given in Figure 5.

## Results and discussion

Plots of data from Table 1 in linear and logarithmic scale for $h$ againts $t$ are given in Figure 6 (left) and (right), respectively.

From Figure 6 (right) value of gradient $m$ from the regression line can be obtained. This value and Equation (11) will give

$$\eta = \frac{1}{(-m)}\left(\frac{\pi R_o^{\,4}\rho g}{8L_o A_i}\right) = \frac{6.361\times 10^{-5}}{(-m)}. \quad (12)$$

Table 1. Experiment data.

| $h$ (cm) | $t$ (s) |
|---|---|
| 22 | 0 |
| 21 | 5 |
| 20 | 9 |
| 19 | 13 |
| 18 | 26 |
| 17 | 34 |
| 16 | 41 |
| 15 | 50 |
| 14 | 56 |
| 13 | 65 |
| 12 | 72 |
| 11 | 78 |
| 10 | 91 |
| 9 | 104 |
| 8 | 115 |
| 7 | 130 |
| 6 | 139 |
| 5 | 155 |
| 4 | 169 |
| 3 | 188 |
| 2 | 215 |
| 1 | 248 |
| 0 | 313 |

Using parameters from the experiment it is found that $\eta$ has value about between 0.374 – 0.707 cP. Two gradient values are obtained, since the results are not too linear, even in logarithmic scale. It is quite good results considering a very rough approximation used in deriving theory for the experiment.

From references it can found that $\eta_{water} \approx 1 \times 10^{-2}$ N·s/m$^2$ or 1 cP at about 20 °C [3] or 0.894 cP at 25 °C [9]. At room temperature it was obtained 0.868 cP, 0.707 cP, and 0.782 cP for buoyant ball experiment, falling ball experiment in Fisika Dasar Lab, and Haake Falling Ball Viscomenter (Type C, Thermo Electron Co.) in Kimia Fisika Lab, respectively [7].

## Conclusion

Using a very simple hand-made viscometer, water viscosity can be measured with obtained value is still in the same order as it is measured by better or standard viscometer, which is also confirmed by a recent result [10]. Further formula simplification is still needed for better use in high school, and for university student assumption to produce Equation (6) can be still debatable. Extension of this work is already implemented in Kompetisi Sains Madrasah 2013 in Malang, Indonesia, 5 - 9 November 2013.

## Acknowledgements

This work is partially supported by RIK ITB 2013 (contract number 248/I.1.C01/PL/2013).

## Referensi

[1] D. Halliday, R. Resnick, and J. Walker, “Fundamentals of Physics”, John Wiley and Sons (Asia), Hoboken, 8th, Extended, Student Edition., 2008, pp. 371-376.
[2] P. M. Fishbane, S. Gasiorowicz, and S. T. Thornton, “Physics for Scientists and Engineers”, Prentice Hall, Upper Saddle River, 2nd, Extended Edition, 1996, pp. 452-453.
[3] W. E. Gettys, F. J. Keller, and M. J. Skove, “Physics Classical and Modern”, McGraw-Hill Book, New York, International Edition, 1989, p. 343-344.
[4] D. Halliday and R. Resnick, “Fisika”, Erlangga, Jakarta, Jilid 1, Edisi 3, 1985, pp. 601-602.
[5] P. A. Tippler, “Fisika untuk Sains dan Teknik”, Erlangga, Jakarta, Jilid 1, Edisi 3, Cetakan 1, 1998, pp. 404-405.
[6] A. F. Abbott, “Ordinary Level Physics”, Heinemann Educational Books, London, 4th Edition, 1984, pp. 146-149.
[7] M. N. Tajuddin, “Eksperimen Bola Bergerak Mengapung di dalam Pipa untuk Menentukan Viskositas Fluida menggunakan Alat Bantu Kamera Digital”, Tesis Magister, Institut Teknologi Bandung, Indonesia, 2009.
[8] A. Soedradjat, “Mekanika-Fluida & Hidrolika”, Nova, 1983, pp. 12-13.
[9] V. L. Streeter dan E. B. Wylie, “Mekanika Fluida”, Erlangga, Jakarta, 1986, p. 175.
[10] L. N. Qomariyatuzzamzami dan S. F. Husen, "Komentar pada 'A Simple Viscometer for High School and First Years Undergraduate Program Students: Theory and Experiment'", Jurnal Pengajaran Fisika Sekolah Menengah 6 (1), 1-3 (2014).

Sparisoma Viridi*
Nuclear Physics and Biophysics
Institut Teknologi Bandung
dudung@fi.itb.ac.id

Sidik Permana
Nuclear Physics and Biophysics
Institut Teknologi Bandung
psidik@fi.itb.ac.id

Wahyu Srigutomo
Earth Physics and Complex System
Institut Teknologi Bandung
wahyu@fi.itb.ac.id

Anggie Susilawati
Physics Department
Universitas Padjajaran
anggie.susilawati@phys.unpad.ac.id

Acep Purqon
Earth Physics and Complex System
Institut Teknologi Bandung
acep@fi.itb.ac.id

*Corresponding author

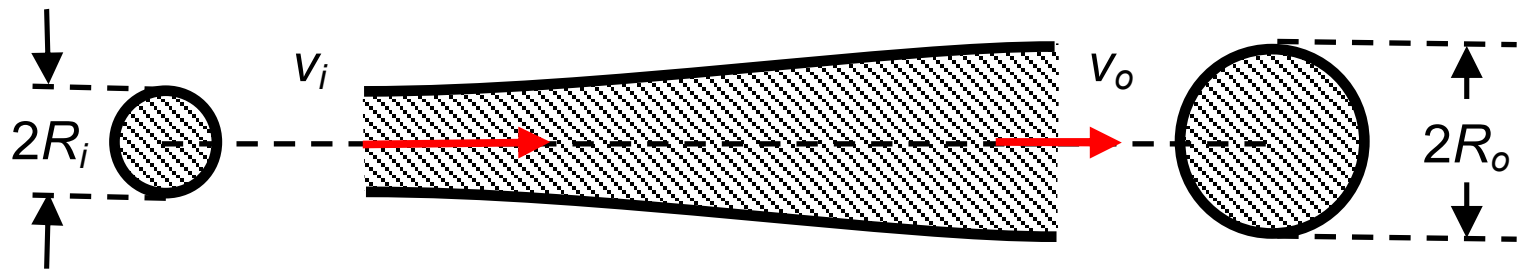

Figure 1. A circular pipe with inlet radius $R_i$ and outlet radius $R_o$.

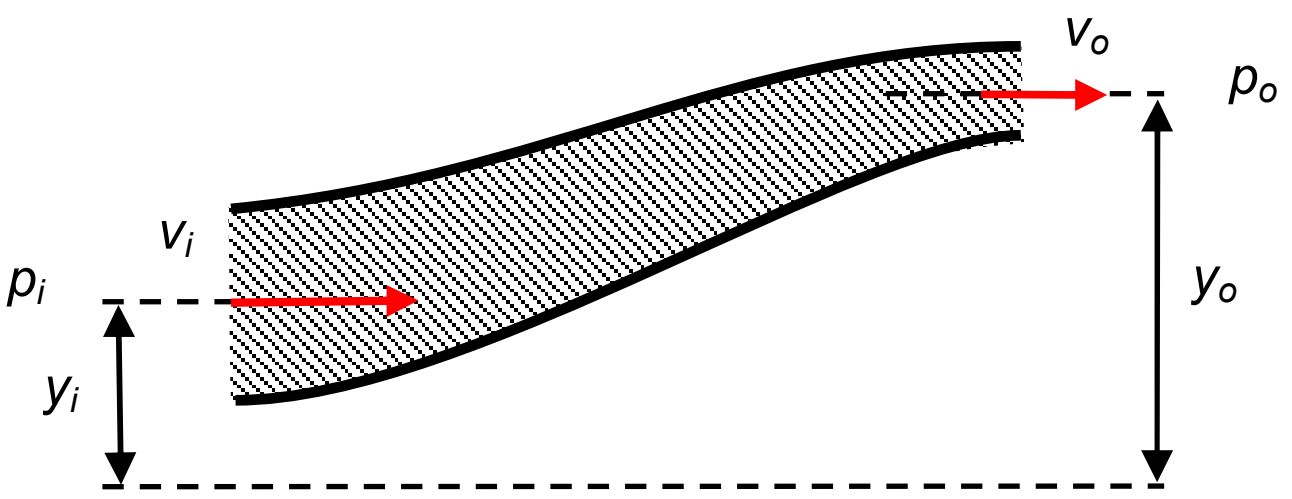

Figure 2. Bernoulli’s equation relates inlet (index $i$) and outlet (index $o$) physical parameters.

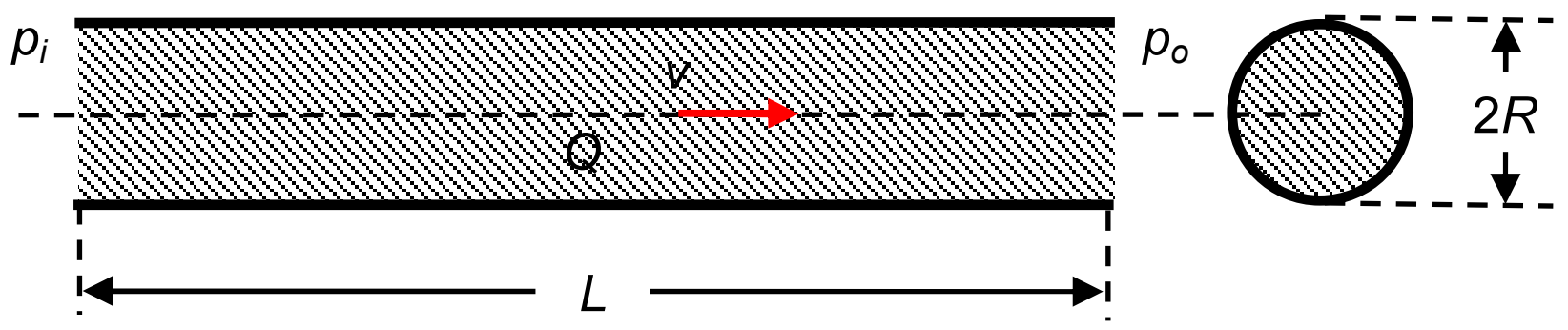

Figure 3. Poiseulli flow through a horizontal pipe, where $\Delta p = p_i - p_o$ is pressure difference between inlet and outlet.

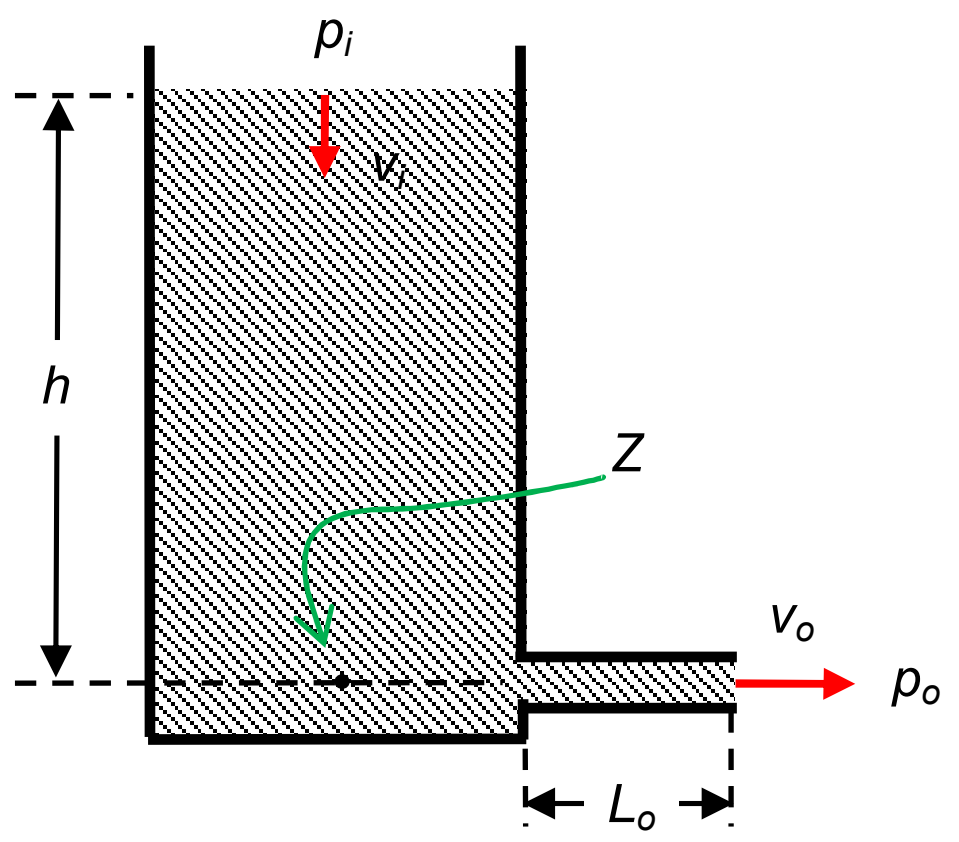

Figure 4. System consists of fluid container with radius $R_i$ and small outlet pipe with radius $R_o$ and length $L_o$, a point $Z$ is defined near the bottom of fluid container and vectically aligned with joint point of the container and the small pipe.

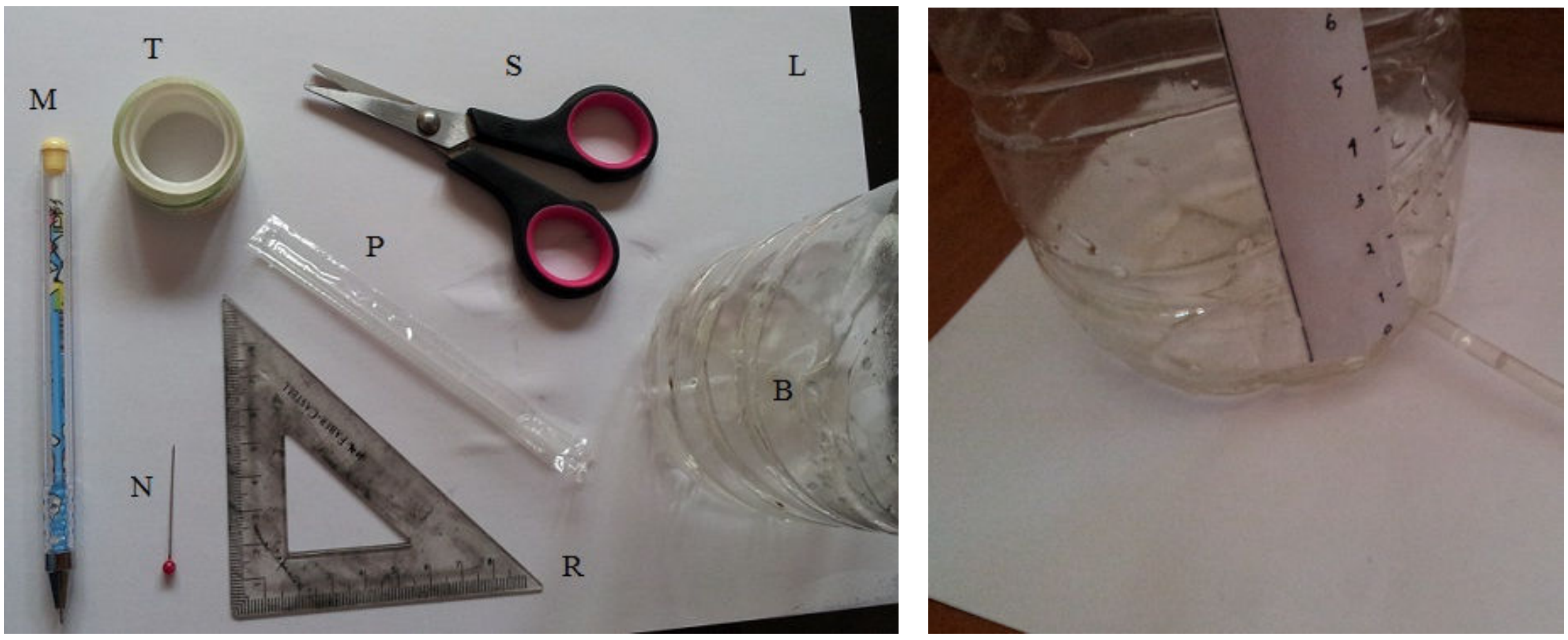

Figure 5. Left: Components to construct simple viscometer are marker (M), transparent tape (T), ruler (R), small drink straw as the small pipe (P), scissor (S), 2-liter mineral water bottle as the fluid container (B), and a piece of paper as label (L). Right: the system after assembled.

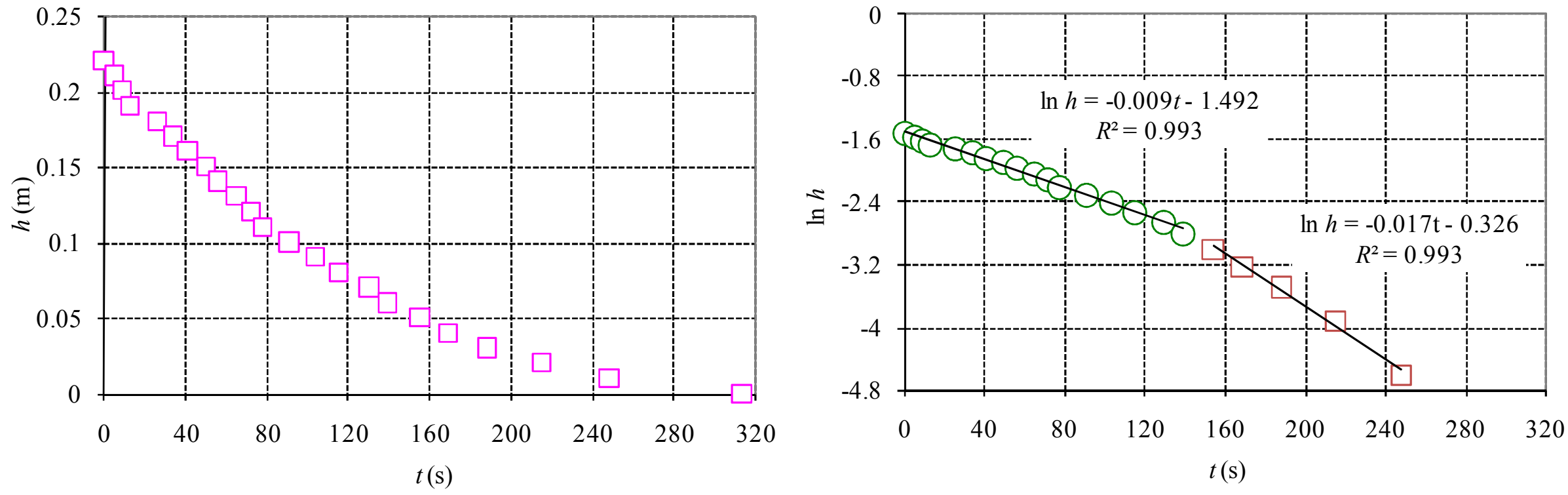

Figure 6. Experiment result for $h$ (left) and $\ln h$ (right) as function of time $t$, where $h$ is represented in m.